\documentclass[twocolumn,showpacs,preprintnumbers,amsmath,amssymb]{revtex4-1}

\usepackage{graphicx}
\usepackage{dcolumn}
\usepackage{bm}
\usepackage{graphics}
\usepackage{subfigure}
\usepackage{graphicx}
\usepackage{epsfig}
\usepackage{epstopdf}

\allowdisplaybreaks

\begin{document}

\title{Interacting Dirac liquid in three-dimensional semimetals}

\author{Johannes Hofmann}

\email{hofmann@umd.edu}

\author{Edwin Barnes}

\author{S. Das Sarma}

\affiliation{Condensed Matter Theory Center and Joint Quantum Institute, Department of Physics, University of Maryland, College Park, Maryland 20742-4111 USA}

\date{\today}

\begin{abstract}
We study theoretically the properties of the interacting Dirac liquid, a novel three-dimensional many-body system which was recently experimentally realized and in which the electrons have a chiral linear relativistic dispersion and a mutual Coulomb interaction. We find that the ``intrinsic" Dirac liquid, where the Fermi energy lies exactly at the nodes of the band dispersion, displays unusual Fermi liquid properties similar to graphene, whereas the ``extrinsic" system with finite detuning or doping behaves as a standard Landau Fermi liquid. We present analytical and numerical results for the self-energy and spectral function based on both Hartree-Fock and the random phase approximation (RPA) theories and compute the quasiparticle lifetime, residue, and renormalized Fermi velocity of the extrinsic Dirac liquid. A full numerical calculation of the extrinsic RPA spectral function indicates that the Fermi liquid description breaks down for large-energy excitations. Furthermore, we find an additional plasmaron quasiparticle sideband in the spectral function which is discontinuous around the Fermi energy. Our predictions should be observable in ARPES and STM measurements.
\end{abstract}

\pacs{71.55.Ak,71.10.Ay,71.45.Gm}
\maketitle

The low-energy excitations of metals or semiconductors are usually well described by a set of parabolic particle and hole bands characterized by an energy offset and an effective mass. The resulting system --- the electron liquid --- is one of the cornerstones of solid state theory~\cite{nozieres97,pines99,giuliani05}. An old question, dating back to the works of Herring~\cite{herring37} and Abrikosov and Beneslavski\u{i}~\cite{abrikosov70}, is whether the band structure of a solid can support more exotic structures where the valence and conduction band touch only at certain points in the Brillouin zone at the Fermi level. The system then forms a semimetallic phase where particle and hole bands are not gapped but at the same time no extended Fermi surface is present. Near the band-touching point, the dispersion takes a chiral relativistic linear form $\varepsilon(p) = \pm v_F p$, where $v_F$ is the Fermi velocity. Indeed, it turns out that such Dirac semimetallic phases can exist generically (i.e., without an artificial fine-tuning, for example, of lattice hopping parameters or spin-orbit coupling strength) and are protected by the crystal symmetry~\cite{young12,yang14}. In addition, if time reversal or inversion symmetry is broken, the Dirac cones are nondegenerate and possess quantum anomalous transport properties~\cite{abrikosov98,yang11,zyuzin12} and topologically protected Fermi arc surface states~\cite{wan11}. Very recently, three-dimensional Dirac semimetals (where the Weyl cones are doubly degenerate) have been realized in Na${}_3$Bi~\cite{liu14,xu15}, Cd${}_3$As${}_2$~\cite{neupane14,borisenko14,liu14b,jeon14,liang15}, and a Weyl semimetal in TaAs~\cite{xu15b}, and the linear dispersion was demonstrated by angle-resolved photoemission spectroscopy (ARPES) measurements of the Dirac cones. These experiments realize a novel fundamental interacting many-body system, which we shall refer to as the Dirac liquid, where the kinetic term has a linear relativistic structure and the electron-electron interaction is the usual $1/r$ nonrelativistic Coulomb interaction. 

In this work, we characterize this Dirac liquid by calculating its quasiparticle properties and spectral function. We compute the quasiparticle lifetime, residue, and renormalized Fermi velocity analytically using Hartree-Fock and the RPA, revealing a standard Fermi liquid behavior at finite carrier density (i.e., doped) and a strange marginal Fermi liquid phase at zero density (i.e., undoped). We calculate the spectral function numerically at finite density and find that the quasiparticle peak is accompanied by sidebands corresponding to plasmaron modes. Our results further indicate a breakdown of the quasiparticle picture away from the Fermi surface as is evident from a significant broadening of spectral weight even for moderate interactions strengths.

We note that related work on the interacting spectral function of the doped 2D material graphene~\cite{polini08,hwang08} has been experimentally verified~\cite{bostwick10,walter11} and considerably enhanced the understanding of this material. It turns out that the spectral function of an extrinsic 3D Dirac liquid has a different structure from the 2D graphene case owing to the different nature of plasmon excitations: while the plasmon mode is not gapped in 2D, implying a plasmaron branch that wraps around the Fermi surface, the gap in the 3D plasmon excitation is manifested in the spectral function as a discontinuity in the plasmaron branch. In addition, the spectral weight at negative energies is focused at the Dirac node, giving rise to a star-shaped structure, whereas in graphene there is a depletion of spectral weight which assumes a ``plasmaron ring" form.

The effective noninteracting two-band Hamiltonian describing the low-energy excitations at one Weyl node is
\begin{align}
H_0 = v_F {\bf k} \cdot \boldsymbol{\sigma} , 
\end{align}
where $v_F$ is the bare (i.e., single-electron) Fermi velocity, ${\bf k}$ the momentum, and $\sigma$ are Pauli matrices. $H_0$ is diagonalized with energy $\varepsilon_s({\bf k}) = s v_F k$ by the chiral eigenstates (for $k_z\neq 0$)
\begin{align}
|{\bf k}s\rangle = \begin{pmatrix}\cos \vartheta_s/2 \\ s \, e^{i\varphi} \sin \vartheta_s/2\end{pmatrix} ,
\end{align}
where $s=\pm1$ is the chirality. The angles $\varphi = \arctan k_y/k_x$ and $\cos \vartheta_+ = k_z/|{\bf k}|$ indicate the direction of the momentum in polar coordinates, and we define $\vartheta_- = \pi - \vartheta_+$. We allow for an arbitrary number $g$ of Weyl nodes in our theory.
We distinguish between the undoped ``intrinsic" case, where the chemical potential is exactly at the nodes of the Dirac dispersion, and the doped ``extrinsic" case with finite chemical potential detuning. The extrinsic case is generic since the presence of impurities or the migration of surface atoms inevitably shifts the chemical potential away from the Dirac node~\cite{liu14,borisenko14,liu14b,neupane14}. We note that the chemical potential can also be tuned experimentally by surface doping, as is, for example, done in recent experiments on Na${}_3$Bi~\cite{liu14}. Following these pioneering experiments, it is imperative to have a quantitatively predictive theoretical calculation of the spectral properties of Dirac liquids taking into account electron-electron interaction effects.

The electrons have a Coulomb interaction
\begin{align}
V(r)= \frac{e^2}{\kappa r} ,
\end{align}
where $e$ is the electron charge and $\kappa$ denotes the effective dielectric constant of the material. Thus, the Coulomb interaction energy scales as the cube root of the carrier density $n$: $E_C = e^2 n^{1/3}/\kappa$. The kinetic energy for linear dispersion possesses the same density dependence $E_K = v_F n^{1/3}$. The dimensionless density-independent interaction strength ($\hbar = 1$ throughout) is set by the ratio of the average interaction to kinetic energy (i.e. the effective fine-structure constant)
\begin{align}
\alpha &= \frac{E_C}{E_K} = \frac{e^2}{\kappa v_F} .
\end{align}
This is to be contrasted with the electron liquid (where the corresponding coupling constant is universally called $r_s$~\cite{nozieres97}), for which the effective interaction strength $r_s \sim n^{-1/3}/\kappa$ depends on the density, indicating that the perturbative limit corresponds to the high-density regime. 

%++++++++++++++++++++++++++++++++++++++++++++++++++++++++++++++++
\begin{figure}[t!]
\scalebox{0.65}{\includegraphics{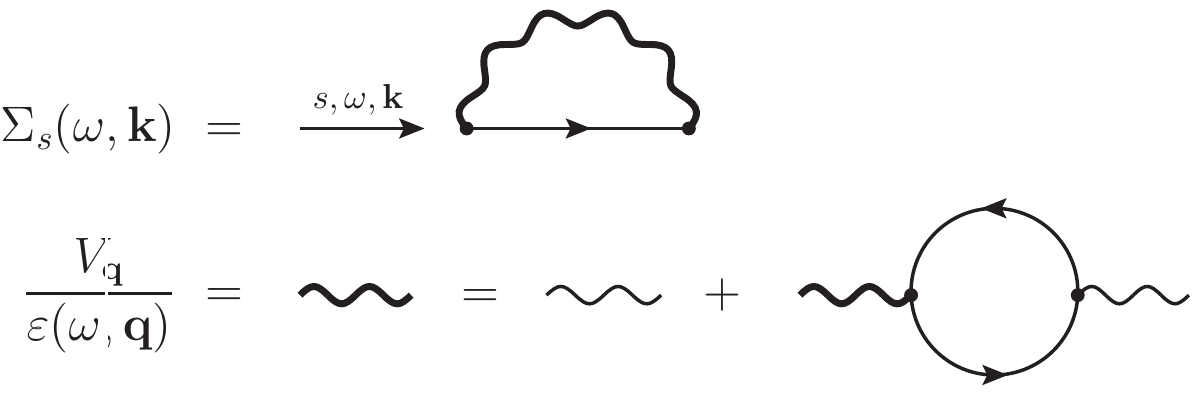}}
\caption{
Diagrammatic representation of the self-energy in the random phase approximation. Thin straight and wiggly lines denote the noninteracting Green's function $G_0$ and the bare Coulomb interaction $V_{\bf q}$, respectively. The dynamically screened RPA interaction (thick wiggly line) is obtained by a summation of the geometric series of Coulomb lines and free polarizability.
}
\label{fig:1}
\end{figure}
%++++++++++++++++++++++++++++++++++++++++++++++++++++++++++++++++

The quasiparticle properties are determined by the self-energy, which is diagonal in a chiral basis:
\begin{align}
\Sigma(\omega, {\bf k}) &= \sum_{s=\pm} \Sigma_s(\omega, {\bf k}) |{\bf k}s\rangle \langle {\bf k}s| .
\end{align}
We compute the self-energy at zero temperature and decompose it into an exchange and correlation part:
\begin{align}
\Sigma_s(\omega, {\bf k}) &= \Sigma_s^{\rm (ex)}(\omega, {\bf k}) + \Sigma_s^{\rm (corr)}(\omega, {\bf k}) .
\end{align}
The exchange part (corresponding to the Hartree-Fock self-energy) contains the leading-order perturbative (i.e., the formal single-loop) interaction correction and is given by
\begin{align}
\Sigma_s^{\rm (ex)}({\bf k}) &= - \frac{1}{V} \sum_{{\bf q}s'} \, \Theta(-\xi_{s'}({\bf q})) V_{{\bf q}-{\bf k}} \, F_{ss'}({\bf k}, {\bf q}) .
\end{align}
Here, the free dispersion is $\xi_s({\bf q}) = v_F(sq-k_F) $ with Fermi momentum $k_F$, $V_{\bf k} = 4\pi e^2/\kappa k^2$ is the Coulomb interaction, and
\begin{align}
F_{ss'}({\bf k}, {\bf q}) = |\langle {\bf k} s | {\bf q}s' \rangle|^2 = \frac{1}{2} (1+ss' \cos \theta)
\end{align}
denotes the overlap of eigenstates of momentum ${\bf q}$ and ${\bf k}$ and chirality $s$ and $s'$. The exchange self-energy can be decomposed into an intrinsic part, which only contains the effects at zero chemical potential, and an extrinsic part, which corrects for the presence of a finite chemical potential:
\begin{align}
\Sigma_s^{\rm (ex)}({\bf k}) &= \Sigma_s^{\rm (int)}(\omega,{\bf k}) + \Sigma_s^{\rm (ext)}(\omega,{\bf k}) , \label{eq:exse}
\end{align}
with
\begin{align}
\Sigma_s^{\rm (int)}({\bf k}) &= - \frac{2\alpha v_Fk_c}{\pi} \left[f\left(\frac{k}{k_c}\right) - s h\left(\frac{k}{k_c}\right)\right] \label{eq:HFint} \\
\Sigma_s^{\rm (ext)}({\bf k}) &= - \frac{2\alpha v_Fk_F}{\pi} \left[f\left(\frac{k}{k_F}\right) + s h\left(\frac{k}{k_F}\right)\right] .
\end{align}
Here, $k_c$ is the ultraviolet momentum scale beyond which we cut off the linear dispersion and
\begin{align}
f(x) &= \frac{1}{4} + \frac{1-x^2}{8x} \ln \left|\frac{1+x}{1-x}\right| , \\
h(x) &= - \frac{1}{12x} + \frac{x}{6} \ln \left|1-\frac{1}{x^2}\right| + \frac{3x^2+1}{24 x^2} \ln \left|\frac{1+x}{1-x}\right| .
\end{align}

Extending this calculation to higher orders in a perturbative calculation in $\alpha$ is not possible without introducing an infrared divergence~\cite{mahan81}. We compute the correlation part in the RPA, or equivalently the $G_0W$ approximation~\cite{hedin65}:
\begin{align}
&\Sigma_s^{\rm (corr)}(i\omega_n,{\bf k}) = - \frac{1}{\beta V} \sum_{{\bf q}s'\Omega_m} \, G_{s'}(i\omega_n+i\Omega_m,{\bf q}') \nonumber \\
&\times V_{\bf q} \biggl[\frac{1}{\varepsilon(i\Omega_m,{\bf q})}-1\biggr] \, F_{ss'}({\bf k}, {\bf q}') , \label{eq:gw}
\end{align}
where ${\bf q}' = {\bf q} + {\bf k}$. Figure~\ref{fig:1} shows the RPA in a Feynman diagram representation. The RPA sums an infinite number of repeated polarization bubble diagrams, leading to a dynamically screened Coulomb interaction with a momentum- and frequency-dependent dielectric function $\varepsilon(i\Omega_m,{\bf q})$ given by
\begin{align}
\varepsilon(i\Omega_m,{\bf q}) &= 1 + V_{\bf q} \Pi(i\Omega_m, {\bf q}) ,
\end{align}
with the polarizability of the noninteracting Dirac liquid
\begin{align}
\Pi(i\omega_n,{\bf q}) &=
- \frac{g}{V} \sum_{{\bf k}ss'} \, \frac{n_F(\xi_s({\bf k})) - n_F(\xi_{s'}({\bf k}'))}{i \omega_n + \xi_s({\bf k}) - \xi_{s'}({\bf k}')} \, F_{ss'}({\bf k}, {\bf k}'),
\end{align}
where $n_F$ is the Fermi-Dirac distribution. The infinite RPA summation removes an infrared divergence stemming from the low-momentum behavior of the polarizability~\cite{mahan81}. Formally, the RPA, which is the leading-order expansion in the dynamically screened Coulomb interaction, is the leading-order term in a systematic large-$g$ expansion, with higher-order corrections suppressed by powers of $1/g$. We note that for many Weyl semimetals, the degeneracy of Weyl nodes is exceptionally large ($g=24$ for pyrochlore iridates~\cite{wan11} or TaAs~\cite{xu15b}), implying that our calculation is essentially exact. In addition, in a previous work~\cite{hofmann14}, we have explicitly computed the next-to-leading order contribution to the RPA in graphene (where $\alpha \leq 2.2$ and $g=4$), which is a two-dimensional Dirac liquid, finding only a very slight correction to the leading-order results for any interaction strength. Therefore, we expect that the RPA provides an excellent description of many-body interaction effects in Dirac liquids.

\section{Intrinsic Dirac liquid}

While the main focus of this work is the generic extrinsic Dirac liquid, for completeness we discuss the intrinsic Dirac liquid to make contact with previous results for graphene and perturbative renormalization group results. To leading order in $\alpha$, the Fermi velocity acquires a logarithmic renormalization which can be extracted from the asymptotic form of the exchange self-energy~\eqref{eq:exse} after subtracting an irrelevant linear divergence~\cite{throckmorton15}. The divergent piece of the intrinsic self-energy~\eqref{eq:HFint} reads
\begin{align}
\Sigma_s^{\rm (int)}({\bf k}) &= s v_F k  \frac{2 \alpha}{3 \pi} \ln \frac{k_c}{k} . \label{eq:HFdiv}
\end{align}
The explicit dependence on $k_c$ can be removed by introducing a counterterm at one-loop order which subtracts the logarithmic divergence~\cite{throckmorton15}. This counterterm introduces a renormalization scale $\mu$, which (in contrast to the cutoff $k_c$) can assume any value. It turns out that this procedure introduces a dependence of the the renormalized Fermi velocity on the renormalization scale $\mu$, which is dictated by a renormalization group equation
\begin{align}
\mu \frac{dv_F}{d\mu} = - \frac{2\alpha}{3\pi} v_F ,
\end{align}
where the coefficient on the right-hand side is set by the logarithmic coefficient of the counterterm and corresponds to the coefficient in Eq.~\eqref{eq:HFdiv}.
Integrating the flow equation for the Fermi velocity with the initial condition $v_F$ at a cutoff scale $k_c$, we obtain
\begin{align}
v_F^* &= v_F \biggl[1 + \frac{2 \alpha}{3 \pi} \ln \frac{k_c}{\mu}\biggr] . \label{eq:exvF}
\end{align}
Here, the renormalization scale $\mu$ is associated with an infrared cutoff to the renormalization group flow, such as (very small) extrinsic density or temperature. Equation~\eqref{eq:exvF} is well known and agrees with previous results~\cite{isobe12,isobe13,rosenstein13,gonzalez14,throckmorton15}. To leading order in the RPA, the polarizability is given by~\cite{abrikosov70}
\begin{align}
\hat{\Pi}^-(\omega,k) &= \frac{k^2}{12 k_F^2} \ln \biggl| \frac{k_c^2}{v_F^2k^2 - \omega^2} \biggr| + i \frac{\pi k^2}{12 k_F^2} \Theta(\omega-v_Fk) , \label{eq:intpol}
\end{align}
where $\hat{\Pi} = \Pi/D_0$ with $D_0 = g k_F^2/2 \pi^2 v_F$ being the noninteracting density of states. Crucially, the logarithmic dependence of the dielectric function on the cutoff $k_c$ induces a renormalization of the electron charge; integrating the one-loop flow equation for the charge~\cite{throckmorton15}, we find that bare and renormalized charge are related via
\begin{align}
\frac{1}{e^2} &= \frac{1}{e_0^2} + \frac{g}{3\pi\kappa v_F} \ln \frac{k_c}{\mu} ,
\end{align}
in agreement with previous work~\cite{isobe12,isobe13,rosenstein13,gonzalez14,throckmorton15}. The charge renormalization here is similar to the renormalization of the electron charge in quantum electrodynamics~\cite{gonzalez14} and does not occur in lower-dimensional Dirac systems such as graphene or the surface of 3D topological insulators. Note that the scale-dependent charge has a spurious divergence at $\Lambda_L = k_c e^{3\pi/g\alpha} > k_c$ which is known as the Landau pole. For an extensive discussion of the renormalization of intrinsic Dirac semimetals, we refer to Refs.~\cite{gonzalez14,throckmorton15}.

In order to characterize the quasiparticle nature of the intrinsic Dirac liquid, we compute the lifetime at small momentum, which is given by the imaginary part of the self-energy. In the on-shell approximation $\omega=v_Fq$, the imaginary part of the intrinsic RPA self-energy vanishes because of phase-space restrictions~\cite{dassarma07}. At zero momentum, however, the self-energy is linear in frequency and given by
\begin{align}
{\rm Im} \Sigma(\omega, {\bf 0}) &=  - \frac{\pi \omega \alpha^2 g}{48} \frac{1}{1+(\pi g\alpha/12)^2},
\end{align}
as is characteristic for a marginal Fermi liquid. Note, however, that the quasiparticle residue does not renormalize to zero~\cite{gonzalez14}, similarly to graphene~\cite{hofmann14}. It turns out that the strange marginal Fermi liquid behavior discussed here breaks down at finite doping.

\section{Extrinsic Dirac liquid}\label{sec:extrinsic}

It is important to note that for an extrinsic system at low temperature, the quasiparticle renormalization is likely overwhelmed by finite density effects: at zero temperature, where the infrared cutoff is set by the density scale, the logarithmic renormalization is only apparent if the density can be tuned extremely close to the Dirac point. In the graphene case, for example, the density must be lowered by two orders of magnitude from the generic extrinsically doped situation to observe a sizable renormalization of the Fermi velocity by a factor of $3$, which requires careful fine-tuning of disorder and doping strength~\cite{elias11}. A similar level of control does not exist for 3D Dirac materials at the present time, and we consider in the following the generic experimentally relevant case of a doped extrinsic Dirac liquid. For this situation, it is sufficient to compute the spectral function and quasiparticle properties for a fixed cutoff, and this is discussed in the remainder of this paper. In addition, there is a possible ambiguity with respect to the choice of the infrared-scale renormalization scale that acts as a cutoff to the renormalization group flow. For an extrinsic system at zero temperature, the full spectral function introduces three low-energy parameters (the doping density as well as the energy and momentum of the excitation), and it is a priori not clear which scale cuts off the renormalization group flow. This problem can be resolved only in few selected cases~\cite{hofmann15}. We emphasize that keeping an explicit cutoff for the finite-density calculation corresponds to the same level of approximation as in graphene~\cite{polini08,hwang08}, for which the RPA agrees very well with experimental measurements~\cite{bostwick10,walter11}. The theory with an explicit finite cutoff is still quantitatively predictive where the cutoff appears as an additional parameter to describe the system (it can be associated, for example, with the scale beyond which the Dirac dispersion is no longer linear). The only thing to bear in mind is that the system is no longer described solely in terms of low-energy (i.e., much smaller than the cutoff) quantities.

We consider first the quasiparticle lifetime in the presence of an extended Fermi surface using the on-shell approximation for the quasiparticle dispersion $\omega=v_F(k - k_F)> 0$. We write the correlation self-energy as the sum of a line and a residue term:
\begin{align}
\Sigma_s^{\rm (corr)}(\omega,{\bf k}) &= \Sigma_s^{\rm (line)}(\omega,{\bf k}) + \Sigma_s^{\rm (res)}(\omega,{\bf k}) \label{eq:lineres} \\
\Sigma_s^{\rm (line)}(\omega,{\bf k}) &= - \frac{1}{V} \sum_{{\bf q}s'} \int \frac{d\Omega}{2\pi} \, G_{s'}(\omega+i\Omega,{\bf q}') \nonumber \\
&\times \biggl[\frac{V_{\bf q}}{\varepsilon(i\Omega,{\bf q})} - 1\biggr] \, F_{ss'}({\bf k}, {\bf q}') \label{eq:seline} \\
\Sigma_s^{\rm (res)}(\omega,{\bf k}) &= \frac{1}{V} \sum_{{\bf q}s'} \, [\Theta(\omega-\xi_{s'}({\bf q}')) - \Theta(-\xi_{s'}({\bf q}'))] \nonumber \\
&\times \biggl[\frac{V_{\bf q}}{\varepsilon(\xi_{s'}({\bf q}')-\omega,{\bf q})}-1\biggr] \, F_{ss'}({\bf k}, {\bf q}') \label{eq:seresidue} ,
\end{align}
where the line contribution is the self-energy~\eqref{eq:gw} with the imaginary frequency replaced by its analytic continuation. The residue term modifies this expression to yield the correct analytic continuation of the retarded self-energy. Only the residue term contributes to the imaginary part and thus the quasiparticle lifetime is given by
\begin{align}
&\frac{1}{\tau(k)} = - 2 {\rm Im} \Sigma_+(\xi_+(k),{\bf k}) = - \frac{1}{(2\pi)^d} \int_{k_F}^k dq\, q^{2}  \nonumber \\
&\times\int d\Omega_d \, (1 + \cos \theta) {\rm Im} \biggl[ \frac{V_{{\bf q}-{\bf k}}}{\varepsilon(v_F (q-k), {\bf q} - {\bf k})} \biggr] . \label{eq:lifetime}
\end{align}
Since we consider low-energy excitations above the Fermi surface, the dielectric function~\cite{lv13} can be expanded in powers of $(k-q)/k_F$; i.e., small frequency density excitations determine the lifetime of the quasiparticle. Furthermore, the Coulomb matrix element implies that the integrand is strongly peaked for forward-scattering processes with $k \approx q$. In this limit, we can use
\begin{align}
\lim_{x\to 0} \lim_{\omega\to 0} {\rm Im} \biggl[ \frac{\varepsilon^{-1}(\omega, x)}{\omega} \biggr] &= - \frac{\pi q}{2 g \alpha v_F k_F^2} .
\end{align}
% ${\rm Im} \, \varepsilon^{-1}(\omega,q) = \pi q \omega/2 g \alpha v_F k_F^2$.
Equation~\eqref{eq:lifetime} becomes
\begin{align}
\frac{1}{\tau(k)} &= - \frac{\pi v_F k_F}{4 g} \int_{1}^x dy \, y^2 (y-x) \nonumber \\
&\qquad \times \int_{-1}^1 d\chi\, \frac{1 + \chi}{\sqrt{x^2+y^2-2xy\chi}} ,
\end{align}
where we set $x=k/k_F$. Performing the integrals, we obtain the standard Fermi liquid expression for the quasiparticle lifetime
\begin{align}
\frac{1}{\tau(k)} &= \frac{\pi}{3 g} \frac{\xi_k^2}{\varepsilon_F} ,
\end{align}
which is quadratic in the excitation energy $\xi_k = v_F (k-k_F)$. Hence, the system behaves as a Fermi liquid with a discontinuity $Z$ in the occupation number at the Fermi surface. 

The quasiparticle residue $Z$ is related to the derivative of the self-energy as $Z^{-1} = 1 - A$, where $A = \lim_{\omega\to 0, k\to k_F} \tfrac{\partial}{\partial\omega}{\rm Re} \Sigma(\omega,{\bf k})$. We compute the self-energy derivative using the line and residue decomposition~\eqref{eq:lineres}. When integrating the line part by parts in $\Omega$, it turns out that the integral boundary term cancels with the residue contribution. Details of the calculation are relegated to the Appendix. It remains to compute
\begin{align}
A &= - \lim_{k\to k_F} {\rm Im} \, \sum_{s'} \int_0^\infty dq \, q^{d-1} \int \frac{d\Omega_d}{(2\pi)^d} \int_0^\infty \frac{d\Omega}{\pi} \nonumber \\
&\times G_{s'}(i \Omega,{\bf q}')  F_{s'}({\bf k}, {\bf q}')  \frac{V_{{\bf q}}}{\varepsilon^2(i\Omega,{\bf q})} \, \frac{\partial \varepsilon(i\Omega,{\bf q})}{\partial \Omega} . \label{eq:defA}
\end{align}
In the weak-interaction small-$\alpha$ limit, the integrand is concentrated in the region of small $k$ and only the intraband excitations with $s'=1$ contribute to the residue. This yields the result
\begin{align}
A &= \frac{\alpha}{\pi^2} \, \int_0^{\pi/2}dz\ln(1-z\cot z) = -1.067 \, \frac{\alpha}{\pi}. \label{eq:A}
\end{align}

The renormalized Fermi velocity is computed in a similar way. It is defined as $v_F^*/v_F = (1+B)/(1-A)$, where $A$ is given in Eq.~\eqref{eq:defA} and $B$ denotes the derivative of ${\rm Re} \Sigma$ with respect to momentum evaluated at zero energy at the Fermi momentum: $B = \tfrac{\partial}{\partial q} {\rm Re} \Sigma_s(\omega, {\bf q}) \bigr|_{\omega=0,q=k_F}$. The calculation proceeds in the same way as for the quasiparticle residue and gives the result for small $\alpha$:
\begin{align}
\frac{v_F^*}{v_F} &= 1 - \frac{\alpha}{2 \pi} \biggl[\ln \frac{g \alpha}{2 \pi} + 2\biggr] + \frac{2 \alpha}{3 \pi} \ln \frac{k_c}{k_F} .
\end{align}
The divergent part agrees with the exchange part~\eqref{eq:exvF} and~\eqref{eq:HFdiv}. We see that in addition to the cutoff-dependent term, there is an additional finite contribution due to intraband interactions.

\section{Spectral function}

%++++++++++++++++++++++++++++++++++++++++++++++++++++++++++++++++
\begin{figure}[t]
\scalebox{0.75}{\includegraphics{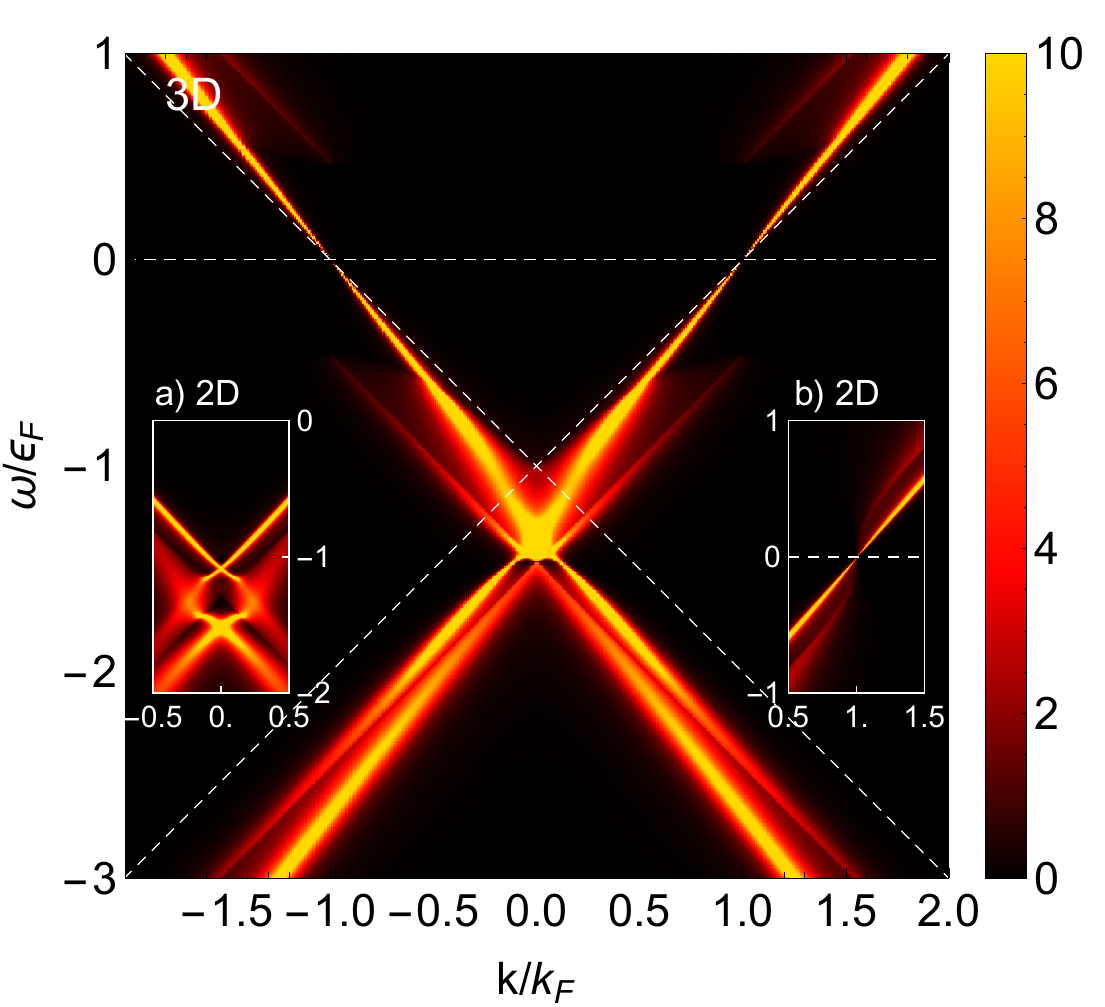}}
\caption{
(Color online) Full spectral function at $\alpha = 0.15$. Clearly visible are the plasmaron sidebands, which have a discontinuous jump across the Fermi momentum. While around the Fermi surface we find a renormalized linear quasiparticle branch, this description breaks down at small momentum and negative frequency, where the quasiparticle cone hybridizes with the plasmaron branch. For comparison, we indicate the peak position of the noninteracting spectral function by white dashed lines. The increased slope of the quasiparticle band is a sign of the Fermi velocity renormalization. The insets show aspects of a typical 2D Dirac material spectral function with $\alpha=0.5$: (a) plasmaron-ring structure near the Dirac point, (b) the plasmaron branch wraps around the Fermi surface, reflecting the ungapped $\sqrt{q}$-2D plasmon mode.
}
\label{fig:spectralfunction}
\end{figure}
%++++++++++++++++++++++++++++++++++++++++++++++++++++++++++++++++

%++++++++++++++++++++++++++++++++++++++++++++++++++++++++++++++++
\begin{figure}[t]
\scalebox{0.6}{\includegraphics{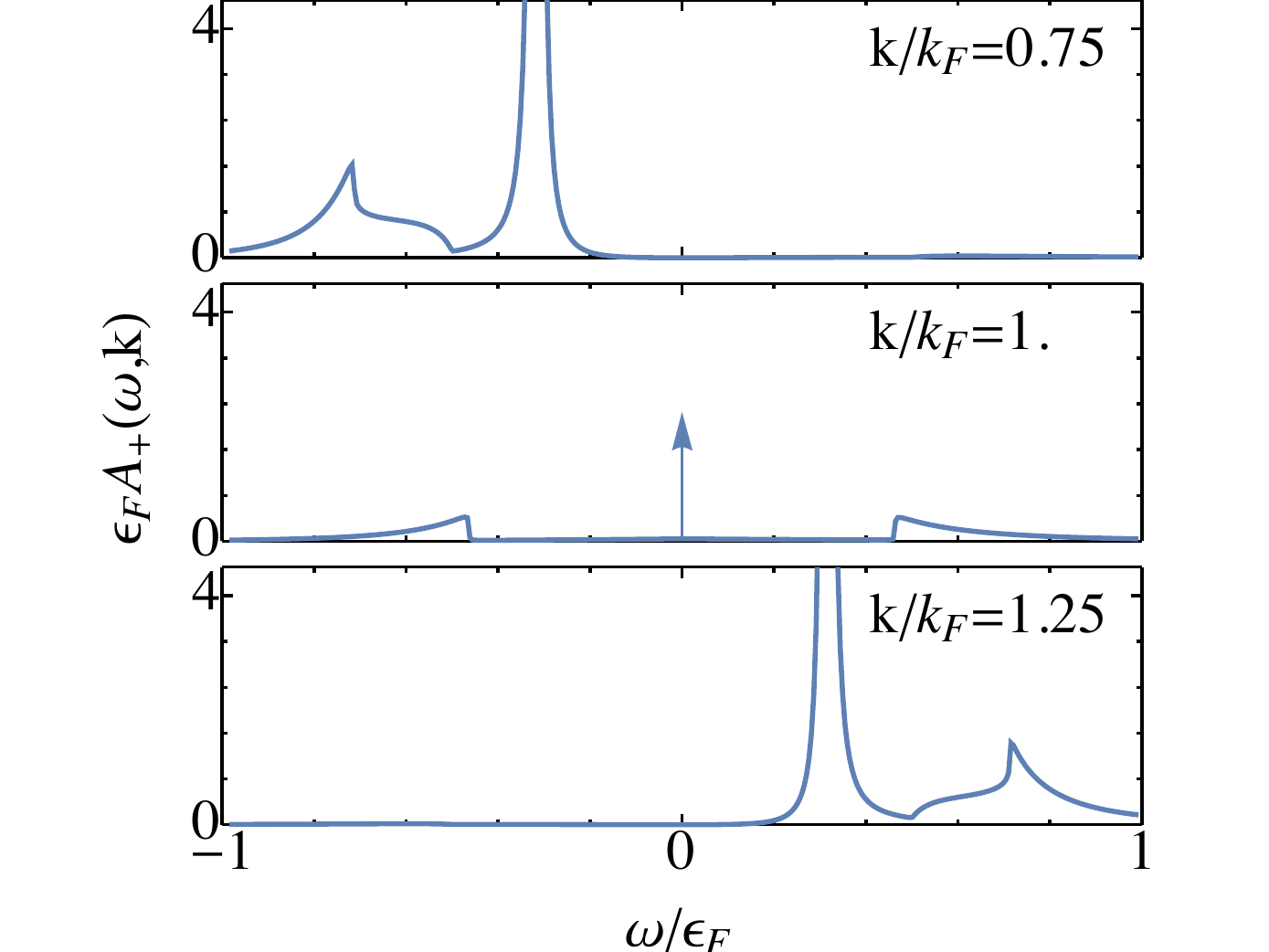}}
\caption{
(Color online) Spectral function at fixed momentum $k/k_F=0.75,1,$ and $1.25$. There is a dominant peak associated with the quasiparticle and a plasmaron side peak. At $k_F$, the quasiparticle peak is a delta function, and the plasmaron excitations are gapped.
}
\label{fig:sliceA}
\end{figure}
%++++++++++++++++++++++++++++++++++++++++++++++++++++++++++++++++

\begin{figure}[t]
\subfigure[]{\scalebox{0.75}{\raisebox{-0.0cm}{\includegraphics{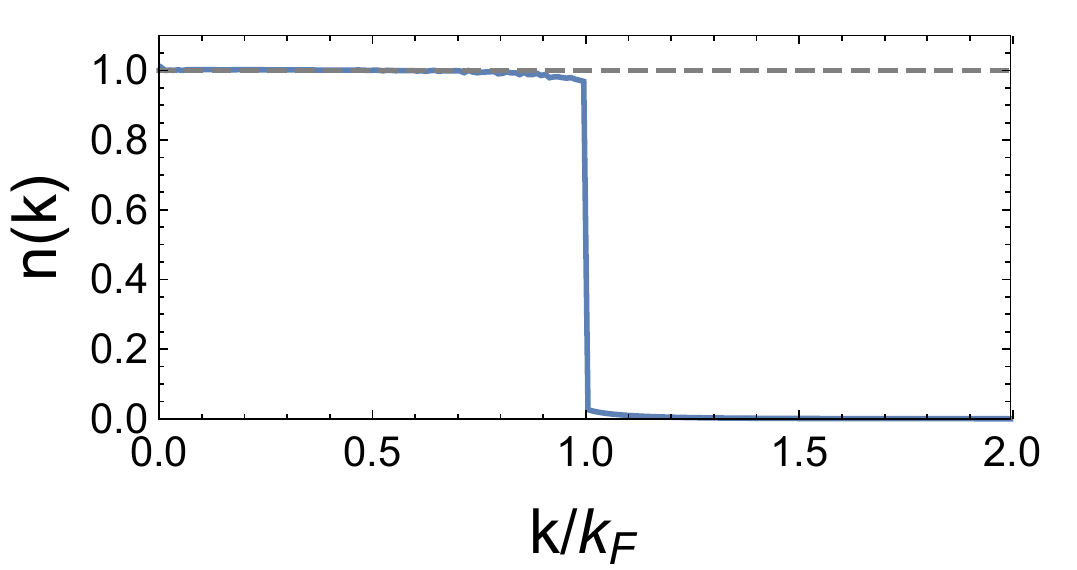}}} \label{fig:occ}}\\
\subfigure[]{\scalebox{0.75}{\raisebox{-0.0cm}{\includegraphics{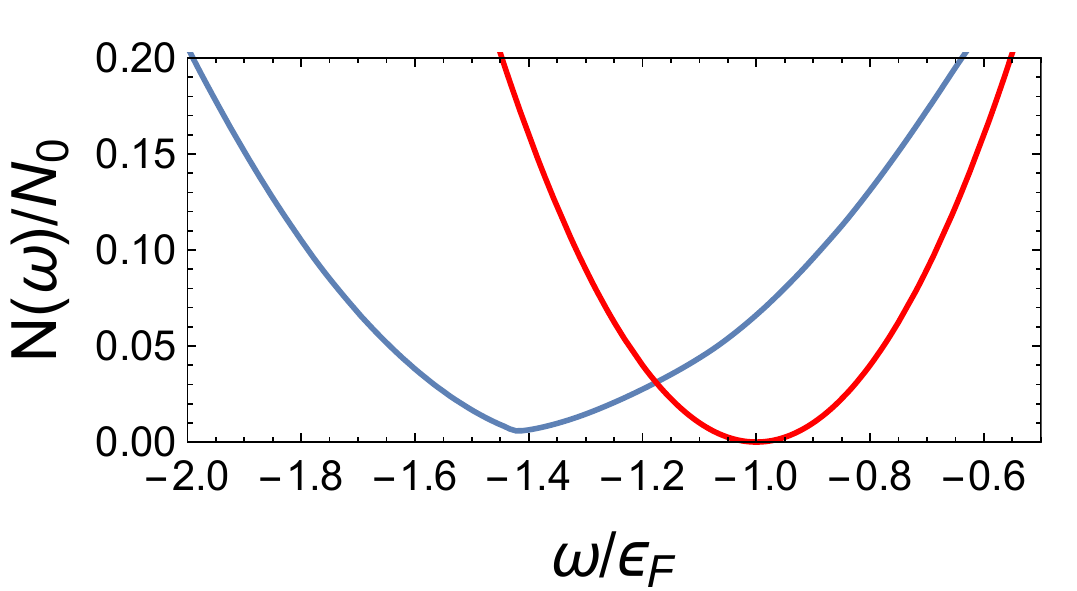}}} \label{fig:dos}}
\caption{
(Color online) (a) Occupation number and (b) density of states. The blue line denotes the interacting result and the red line denotes the noninteracting DOS.
}
\end{figure}

Having analytically calculated the quasiparticle properties, we proceed to present numerical results for the full spectral function. The full spectral function is given by
\begin{align}
A(\omega, {\bf q})  =  \sum_{s=\pm} A_s(\omega, {\bf q}) ,
\end{align}
where
\begin{align}
A_s(\omega, {\bf k}) &=  \frac{1}{\pi} \frac{{\rm Im} \Sigma_s(\omega, {\bf k})}{(\omega - \xi_s({\bf k}) - {\rm Re} \Sigma_s(\omega, {\bf k}))^2 + ({\rm Im} \Sigma_s(\omega, {\bf k}))^2} .
\end{align}
Figure~\ref{fig:spectralfunction} shows the spectral function as computed numerically from Eqs.~\eqref{eq:seline} and~\eqref{eq:seresidue} at an interaction strength $\alpha=0.15$ and with a cutoff $k_c/k_F = 100$. As is standard for the $G_0W$ approximation, we compute the real part of the self-energy relative to the chemical potential at the Fermi surface. As a consistency check of our calculation, we verify the normalization of the spectral function to within less than one percent. Beside the quasiparticle branch discussed in the previous sections, the coupling of electrons to the plasmon density modes induces an additional quasiparticle excitation, the plasmaron~\cite{lundqvist67}, which forms sidebands to the quasiparticle dispersion. These plasmaron satellite peaks are clearly visible in Fig.~\ref{fig:sliceA} which shows the spectral function as a function of frequency at fixed momentum $k/k_F=0.75,1$, and $1.25$. For a momentum smaller than $k_F$, the plasmaron band is below the quasiparticle branch, whereas above $k_F$, it lies above it. Right at the Fermi momentum, the quasiparticle peak has infinite lifetime and zero width as indicated by the arrow. Since the plasmon mode is gapped in three dimensions with a dispersion of $\omega_p(q) = \omega_0 + {\cal O}(q^2)$, where $\omega_0 > 0$, low-energy excitations around zero frequency do not couple to the plasmon modes and, hence, there is a discontinuous jump in the plasmaron branch across the Fermi surface. Note that this is in sharp contrast to 2D Dirac materials such as graphene, where an ungapped plasmon mode implies a plasmaron mode at any infinitesimal excitation energy [cf. inset (b) of Fig.~\ref{fig:spectralfunction}].
Note that in the 2D graphene case, while RPA finds a plasmaron quasiparticle pole, a numerical GW cumulant expansion reports a plasmaron branch solely due to the aggregation of spectral weight~\cite{lischner13}, at variance with experimental findings~\cite{bostwick10,walter11}. The Fermi liquid nature of our system is also apparent in the occupation number shown in Fig.~\ref{fig:occ}, which clearly displays the discontinuity of size $Z$ at the Fermi momentum. We note that for excitations at small momentum or negative chirality, the quasiparticle description breaks down. In addition, we observe strong interaction effects on the density of states [Fig.~\ref{fig:dos}], the minimum of which is shifted with respect to the minimum of the noninteracting DOS which is located at $\omega=-\varepsilon_F$. At high frequency, the DOS retains its noninteracting quadratic shape with a renormalized slope $\sim v_F/v_F^*$.

%++++++++++++++++++++++++++++++++++++++++++++++++++++++++++++++++
%++++++++++++++++++++++++++++++++++++++++++++++++++++++++++++++++ 

\section{Summary}

In summary, we provide a comprehensive theory for the quasiparticle properties of the Dirac liquid in the random phase approximation. While we found that the intrinsic Dirac liquid displays a strange marginal Fermi liquid behavior with an imaginary part that is linear in frequency, the experimentally relevant extrinsic Dirac liquid behaves as a Fermi liquid. We have computed the Fermi liquid properties---lifetime, residue, and renormalized Fermi velocity---both analytically in the on-shell approximation as well as numerically in a full calculation of the spectral function. The numerical calculation reveals the limit of the Fermi liquid description away from the Fermi surface and indicates the existence of an additional quasiparticle excitation, the plasmaron, which arises from a coupling to the plasmon density modes. Our predictions can be directly tested in spectroscopic measurements such as ARPES and STM.

This work is supported by LPS-MPO-CMTC.

\appendix

\section{Quasiparticle properties}

In this Appendix, we give details of the calculation of quasiparticle residue and Fermi velocity presented in Sec.~\ref{sec:extrinsic}.

\subsection{Quasiparticle residue}

In the following, we evaluate the quasiparticle weight for excitations close to the Fermi surface. To this end, we compute
\begin{align}
A &= - \lim_{\omega \to 0, q\to k_F} \frac{\partial}{\partial \omega} {\rm Re} \frac{1}{V} \sum_{s'} \sum_{\bf k} \nonumber \\
&\int \frac{d\Omega}{2\pi} \, G_{s'}(i\omega_n+i\Omega,{\bf k}') \frac{V_{\bf k}}{\varepsilon(i\Omega,{\bf k})} \, F_{s'}({\bf q}, {\bf k}') \biggl|_{i\omega_n \to \omega + i0}.
\end{align}
This expression can be split in a line and a residue term~\cite{mahan81}:
\begin{align}
A &= A^{\rm (line)} + A^{\rm (res)} \\
A^{\rm (line)} &= - \lim_{\omega \to 0, q\to k_F} \frac{1}{V} \sum_{s'} \sum_{\bf k} \int \frac{d\Omega}{2\pi} \, \frac{\partial}{\partial \omega} G_{s'}(\omega+i\Omega,{\bf k}') \nonumber \\
&\qquad \times \frac{V_{\bf k}}{\varepsilon(i\Omega,{\bf k})} \, F_{s'}({\bf q}, {\bf k}') \\
A^{\rm (res)} &= \lim_{q\to k_F} {\rm Re} \, \frac{1}{V} \sum_{s'} \sum_{\bf k} \, \delta(- \xi_{s'}({\bf k}')) \nonumber \\
&\qquad \times \frac{V_{\bf k}}{\varepsilon(\xi_{s'}({\bf k}'),{\bf k})} \, F_{s'}({\bf q}, {\bf k}') \label{eq:Ares} .
\end{align}
Since the static polarizability is real, the last expression~\eqref{eq:Ares} is real and we can omit explicitly taking the real part. The line part can be written as
\begin{align}
A^{\rm (line)} &= \lim_{q\to k_F} {\rm Re} \, \frac{2 i}{V} \sum_{s'} \sum_{\bf k} \int_0^\infty \frac{d\Omega}{2\pi} \, \biggl[\frac{\partial}{\partial \Omega} G_{s'}(i\Omega,{\bf k}')\biggr] \nonumber \\
&\qquad \times \frac{V_{\bf k}}{\varepsilon(i\Omega,{\bf k})} \, F_{s'}({\bf q}, {\bf k}') .
\end{align}
Integrating this expression by parts in $\Omega$~\cite{saraga05,dassarma07}, we obtain
\begin{align}
A^{\rm (line)} &= A^{\rm (bd)} + \lim_{q\to k_F} {\rm Im} \, \frac{1}{V} \sum_{s'} \sum_{\bf k} \int_0^\infty \frac{d\Omega}{\pi} \, G_{s'}(i \Omega,{\bf k}')  \nonumber \\
&\qquad \times F_{s'}({\bf q}, {\bf k}') \, \frac{\partial}{\partial \Omega} \frac{V_{\bf k}}{\varepsilon(i\Omega,{\bf k})}  ,
\end{align}
where the boundary term arises from the lower integration boundary $\Omega=0+$. It reads
\begin{align}
A^{\rm (bd)} &= - \lim_{q\to k_F} {\rm Re} \, \frac{i}{\pi V} \sum_{s'} \sum_{\bf k} \, G_{s'}(i0+,{\bf k}') \nonumber \\
&\qquad \times \frac{V_{\bf k}}{\varepsilon(i0+,{\bf k})} \, F_{s'}({\bf q}, {\bf k}') \nonumber \\
&= - \lim_{q\to k_F} {\rm Re} \, \frac{1}{V} \sum_{s'} \sum_{\bf k} \, \delta(\xi_{s'}({\bf k}')) \frac{V_{\bf k}}{\varepsilon(0,{\bf k})} \, F_{s'}({\bf q}, {\bf k}') ,
\end{align}
and thus cancels with the residue contribution~\eqref{eq:Ares}. It remains to compute the integral given in Eq.~\eqref{eq:defA} of the main text.
%\begin{align}
%A &= - \lim_{q\to k_F} {\rm Im} \, \sum_{s'} \int_0^\infty dk \, k^{d-1} \int \frac{d\Omega_d}{(2\pi)^d} \int_0^\infty \frac{d\Omega}{\pi} \, G_{s'}(i \Omega,{\bf k}')  F_{s'}({\bf q}, {\bf k}')  \frac{V_{{\bf k}}}{\varepsilon^2(i\Omega,{\bf k})} \, \frac{\partial \varepsilon(i\Omega,{\bf k})}{\partial \Omega} 
%\end{align}
We shall evaluate this expression in the weak-interaction small-$\alpha$ limit, in which case the integrand is concentrated in the region of small ${\bf k}$. In the following, we will make use of dimensionless variables and define $x=q/k_F$ and $y=k/k_F$. Since $F_{s'}({\bf x}, {\bf y}') = (1 + s')/2 + {\cal O}((y/x)^2)$, we only take into account the interband contribution with $s'=1$. Furthermore, $[G_{+}(i \Omega,{\bf y}')]^{-1} = i \Omega - y \cos \theta + {\cal O}((y/x)^2)$:
\begin{align}
A &= - \frac{\alpha}{\pi} \, {\rm Im} \, \int_0^\infty dy \int \frac{d\Omega_d}{\Omega_d} \int_0^\infty d\Omega \, \frac{1}{i\Omega - y \cos \theta} \nonumber \\
&\qquad \times \frac{1}{\varepsilon^2(i\Omega,{\bf y})} \frac{\partial \varepsilon(i\Omega,{\bf y})}{\partial \Omega} .
\end{align}
We infer the small-$y$ limit of the dielectric function taking into account only the extrinsic polarizability:
\begin{align}
\hat{\Pi}^+(i\Omega , {\bf x}) &= - \int \frac{d\Omega_d}{\Omega_d} \frac{\cos \theta}{(i \Omega/x) + \cos \theta} \nonumber \\
&= \frac{\Omega}{x} \arctan \dfrac{x}{\Omega} - 1 .
\end{align}
We can remove the $y$ dependence from the Green's function contribution to the integrand by defining the rescaled integration variable $u=\Omega/y$:
\begin{align}
A &= - \frac{g \alpha^2}{\pi} \, {\rm Im} \, \int_0^\infty du \, \frac{\partial a(u)}{\partial u} \int \frac{d\Omega_d}{\Omega_d} \frac{1}{iu - \cos \theta} \nonumber \\
&\qquad \times \int_0^\infty dy \, \frac{y^{d-2}}{(y^{d-1} + g\alpha a(u))^2} \nonumber \\
&= \frac{g \alpha^2}{\pi (d-1)} \, {\rm Im} \, \int_0^\infty du \, \frac{\partial a(u)}{\partial u} \int \frac{d\Omega_d}{\Omega_d} \frac{1}{iu - \cos \theta} \nonumber \\
&\qquad \times  \int_0^\infty dy \, \frac{\partial}{\partial y} \, \frac{1}{y^{d-1} + g\alpha a(u)} \nonumber \\
&= - \frac{\alpha}{\pi (d-1)} \, {\rm Im} \, \int_0^\infty du \, \frac{1}{a(u)} \frac{\partial a(u)}{\partial u} \int \frac{d\Omega_d}{\Omega_d} \frac{1}{iu - \cos \theta} \nonumber \\
&= \frac{\alpha}{\pi} \, \int_0^\infty du \, \frac{1}{a(u)} \frac{\partial a(u)}{\partial u} \frac{1}{2} \arctan \frac{1}{u} ,
\end{align}
where
\begin{align}
a(u) &= u \arctan \frac{1}{u} - 1 .
\end{align}
The $u$ integration can be performed only numerically in 3D. The result is given in Eq.~\eqref{eq:A}.

\subsection{Fermi velocity}

A similar analysis to that for the quasiparticle residue applies. We have
\begin{align}
B &= - \lim_{\omega \to 0, q\to k_F} \frac{\partial}{\partial q} {\rm Re} \frac{1}{V} \sum_{s'} \sum_{\bf k} \int_{-\infty}^\infty \frac{d\Omega}{2\pi} \, G_{s'}(i\omega_n+i\Omega,{\bf k}') \nonumber \\
&\qquad \times \frac{V_{\bf k}}{\varepsilon(i\Omega,{\bf k})} \, F_{s'}({\bf q}, {\bf k}') \biggl|_{i\omega_n \to \omega + i0}.
\end{align}
The derivative of the residue term with respect to the external momentum vanishes. We are left with the line contribution:
\begin{align}
B &= - \lim_{q\to k_F} \frac{1}{V} \sum_{s'} \sum_{\bf k} \int_{-\infty}^\infty \frac{d\Omega}{2\pi} \, \frac{V_{\bf k}}{\varepsilon(i\Omega,{\bf k})} \nonumber \\
&\qquad \times \frac{\partial}{\partial q} \biggl[ G_{s'}(i\Omega,{\bf k}') F_{s'}({\bf q}, {\bf k}') \biggr] .
\end{align}
We split this expression into two parts:
\begin{align}
B &= B_1 + B_2 \\
B_1 &= - \lim_{q\to k_F} {\rm Re} \frac{1}{V} \sum_{s'} \sum_{\bf k} \int_0^\infty \frac{d\Omega}{\pi} \, \frac{V_{\bf k}}{\varepsilon(i\Omega,{\bf k})} \, \frac{\partial k'}{\partial q} \nonumber \\
&\qquad \times \biggl[ \frac{\partial}{\partial k'} G_{s'}(i\Omega,{\bf k}')\biggr] F_{s'}({\bf q}, {\bf k}') \\
B_2 &= - \lim_{q\to k_F} \frac{1}{V} \sum_{s'} \sum_{\bf k} \int_{-\infty}^\infty \frac{d\Omega}{2 \pi} \, \frac{V_{\bf k}}{\varepsilon(i\Omega,{\bf k})} \, G_{s'}(i\Omega,{\bf k}') \nonumber \\
&\qquad \times \biggl[ \frac{\partial}{\partial q} F_{s'}({\bf q}, {\bf k}') \biggr] .
\end{align}
Consider the first term $B_1$. We use
\begin{align}
\frac{\partial}{\partial k'} G_{s'}(i\Omega,{\bf k}') &= i v_F s' \frac{\partial}{\partial \Omega} G_{s'}(i\Omega,{\bf k}') ,
\end{align}
and integrate by parts in $\Omega$. This gives another $\Omega$ integral and a boundary term:
\begin{align}
B_1 &= B_1^{\rm (bd)} + B_1' \\
B_1^{\rm (bd)} &= \lim_{q\to k_F} \frac{1}{V} \sum_{s'} \sum_{\bf k} s' \frac{V_{{\bf k}-{\bf q}}}{\varepsilon(0,{\bf k}-{\bf q})} \, \delta(\mu - \varepsilon_{s'}({\bf k})) \nonumber \\
&\qquad \times F_{s'}({\bf q}, {\bf k}) \cos \theta \\
B_1' &= \lim_{q\to k_F} {\rm Im} \, \frac{1}{V} \sum_{s'} \sum_{\bf k} s' \int_0^\infty \frac{d\Omega}{\pi} \,  G_{s'}(i\Omega,{\bf k}') \nonumber \\
&\qquad \times F_{s'}({\bf q}, {\bf k}') \, \frac{\partial k'}{\partial q} \, \biggl[ \frac{\partial}{\partial \Omega} \frac{V_{\bf k}}{\varepsilon(i\Omega,{\bf k})} \biggr] .
\end{align}
The boundary term is (switching to dimensionless variables)
\begin{align}
B_1^{\rm (bd)} &= \lim_{x,y\to 1} \, \frac{\alpha}{2} \int \frac{d\Omega_d}{\Omega_d} \frac{\cos \theta}{|{\bf y}-{\bf x}|^{d-1} \varepsilon(0,{\bf y}-{\bf x})} (1+\cos\theta) \nonumber \\
&= 
- \frac{\alpha}{4} \biggl(\ln \frac{g \alpha}{4} + 2\biggr) . \label{eq:b1bd}
\end{align}
The integrand of $B_1'$ is equivalent the residue integrand in Eq.~\eqref{eq:Ares} up to the factor of
\begin{align}
\frac{\partial k'}{\partial q} &= \frac{{\bf q} \cdot {\bf k}'}{q k'} = 1 + {\cal O}((k/k_F)^2) .
\end{align}
Thus, for small $\alpha$, the calculation proceeds exactly as for $A$, and we obtain
\begin{align}
B_1' &= 1.677 \, \frac{\alpha}{\pi} .
\end{align}
It remains to compute
\begin{align}
B_2 &= - \lim_{q\to k_F} \frac{1}{V} \sum_{s'} \sum_{\bf k} \int_{-\infty}^\infty \frac{d\Omega}{2 \pi} \, \frac{V_{\bf k}}{\varepsilon(i\Omega,{\bf k})} \, G_{s'}(i\Omega,{\bf k}')  \nonumber \\
&\qquad \times \biggl[ \frac{\partial}{\partial q} F_{s'}({\bf q}, {\bf k}') \biggr] .
\end{align}
Note that
\begin{align}
\frac{\partial}{\partial q} F_{s'}({\bf q}, {\bf k}') &= \frac{s' k^2 \sin^2 \theta}{2 k'^3} 
\end{align}
exactly, which implies that there is no short-range singularity for small ${\bf k}$. To leading order in $\alpha$, we thus consider the static limit $\varepsilon(i\Omega,{\bf k}) = 1$. In dimensionless units:
\begin{align}
&B_2 = - \frac{\alpha}{2} \lim_{x\to 1} \sum_{s'} s' \int_0^\infty dy \int \frac{d\Omega_d}{\Omega_d} \int_{-\infty}^\infty \frac{d\Omega}{2\pi} \, G_{s'}(i\Omega,{\bf y}') \nonumber \\
&\qquad \times \frac{y^2 \sin^2 \theta}{y'^3} \nonumber \\
&= - \frac{\alpha}{2} \sum_{s'} s' \int_0^\ell dy \int \frac{d\Omega_d}{\Omega_d} \, \Theta(-\xi_{s'}({\bf y}')) \nonumber \\
&\qquad \times \frac{y^2 \sin^2 \theta}{[1+y^2-2y\cos\theta]^{3/2}} \nonumber \\
&= B_{2,+} - B_{2,-} ,
\end{align}
with $B_{2,+}$ ($B_{2,-}$) denoting the interband (intraband) contribution, respectively. We have
\begin{align}
B_{2,+} &= \frac{\alpha}{12} \bigl(1 - 4 \ln 2\bigr) \\
B_{2,-} &= - \frac{\alpha}{3} \ln \ell - \frac{\alpha}{9} .
\end{align}
The finite pieces of $B_2$ almost cancel and are small compared to the corresponding finite term in $B_1^{\rm (bd)}$, Eq.~\eqref{eq:b1bd}. Neglecting $B_2$, the result for the Fermi velocity is
\begin{align}
\frac{v_F^*}{v_F} &= \frac{1+B}{1-A} = 1 - \frac{\alpha}{4} \biggl[\ln \frac{g \alpha}{4} + 2\biggr] + \frac{\alpha}{3} \ln \frac{\Lambda}{k_F} .
\end{align}
The divergent piece of the Fermi velocity agrees with the exchange part result and with previous work~\cite{rosenstein13}.

\bibliography{bib}

\end{document}